\documentclass{ws-ijmpa}

\usepackage{amsmath}
\usepackage{amsfonts}
\usepackage{amscd}
\usepackage{epsfig}
\usepackage{amssymb}
\usepackage{tabularx}

\def\w{\omega}

\begin{document}

%\markboth{Gustavo Dotti, Julio Oliva and Ricardo Troncoso}
%{Vacuum solutions with nontrivial boundaries for Einstein-Gauss-Bonnet}

\catchline{}{}{}{}{}
%
%%%%%%%%%%%%%%%%%%%%%%%%%%%%%%%%%%%%%%%%%%%%%%%%%%%%%%%%%%%%%%%%%%%%

\title{Instabilities of naked singularities and black hole interiors in General Relativity}

\author{Gustavo Dotti and Reinaldo J. Gleiser}

\address{Facultad de Matem\'{a}tica, Astronom\'{\i}a y F\'{\i}sica, Universidad
Nacional de C\'{o}rdoba,\\ Ciudad Universitaria, (5000) C\'{o}rdoba, Argentina.}

\author{Jorge Pullin}

\address{Department of
Physics and Astronomy, Louisiana State University, Baton Rouge, LA
70803-4001}

\author{Ignacio F. Ranea-Sandoval and H\'ector Vucetich}

\address{Facultad de Ciencias Astron\'omicas y Geof\'{\i}sicas, Universidad Nacional de La Plata.
 Paseo del Bosque S/N 1900. La Plata, Argentina}

\maketitle

\begin{history}
\received{Day Month Year}
\revised{Day Month Year}
\end{history}

\begin{abstract}
Metrics representing black holes in General Relativity may exhibit
naked singularities for certain values of their parameters. This is  the case for
 super-extremal
($J^2 > M>0$) Kerr  and  super-extremal ($|Q|>M>0$) Reissner-N\"ordstrom spacetimes,
and also for the
negative mass Schwarzschild spacetime. We review our recent work where we show
that these nakedly singular spacetimes are unstable under
linear gravitational perturbations, a result that supports the cosmic censorship
conjecture, and  also  that the inner stationary region beyond the inner horizon
of a Kerr black hole ($J^2 <M$) is linearly unstable.

\end{abstract}

\ccode{PACS numbers: 04.50.+h, 04.20.Jb, 04.90.+e}

Consider the Schwarzschild solution for the Einstein's vacuum equations in the  static, spherically symmetric case,
\begin{equation}\label{chuar}
ds^2 = -\left(1-\frac{2 {M}}{r} \right) dt^2 +
\frac{dr^2}{1-\frac{2 {M}}{r}} + r^2 ( d\theta^2 + \sin^2 \theta d\phi^2).
\end{equation}
Its generalization to the axially symmetric (rotating) case, obtained by Kerr, is
\begin{multline}\label{kerr}
ds^2 = - \frac{(\Delta-a^2 \sin^2 \theta) }{\Sigma} dt^2 -2a\sin^2
\theta \frac{(r^2 + a^2 - \Delta)}{\Sigma}
dt d\phi \\
+ \left[ \frac{(r^2+a^2)^2 - \Delta a^2 \sin^2 \theta}{\Sigma} \right] \sin^2 \theta d\phi^2
+
\frac{\Sigma}{\Delta} dr^2 + \Sigma  d\theta^2 ,
\end{multline}
where $\Sigma = r^2 + a^2 \; \cos ^2 \theta$ and $ \; \Delta = r^2-2Mr + a^2$.
The generalization of (\ref{chuar}) to the spherically symmetric Einstein-Maxwell case is
the Reissner-N\"ordstrom spacetime, with electromagnetic field $F =   \frac{Q}{r^2} \; dt \wedge dr$ and  metric
\begin{equation}\label{rn}
ds^2 = -\left(1-\frac{2 {M}}{r}+\frac{ {Q}^2}{r^2} \right) dt^2 +
\frac{dr^2}{1-\frac{2 {M}}{r}+\frac{ {Q}^2}{r^2} } + r^2 ( d\theta^2 + \sin^2 \theta d\phi^2).
\end{equation}
$a:=J/M$, $Q$ and $M$ above arise as integration constants when solving the second order field equations,
and can therefore  take any value. We know, however, that $Q$ is the charge, $M$ the mass, and $J$ the angular momentum,
and that $|Q| > M (>0)$ in (\ref{rn}), as well as $|J| > M^2$ in (\ref{kerr}), or a negative $M$ in
(\ref{chuar}),  would give a spacetime with a naked singularity. Although not ruled
out by Einstein's equations, it is believed that a nakedly singular stationary spacetime cannot be the
endpoint of the evolution of ``reasonable'' evolving matter. Different technical versions of this assertion
are usually referred to as ``cosmic censorship''. We have addressed this issue through a series of papers \cite{dg,dgp,dgsv,dg2}
that analyze the stability under linear gravitational perturbations of these nakedly singular spacetimes, and
found that they are generically unstable, providing further supporting evidence to the cosmic censorship conjecture.\\

For the $M<0$ Schwarzschild solution, the instability, first analyzed in Ref.5,
 corresponds to the scalar (even) type \cite{ki} of gravitational
perturbations \cite{dg,dg2}. It was  Zerilli \cite{zerilli} who found that, for the black hole case $M>0$,
the scalar type linear perturbations for   the static exterior region $r>2M$ of a Schwarzschild black hole
can be reduced to a two dimensional wave equation
with a space dependent  potential
\begin{equation} \label{z}
\frac{\partial^2 \Psi_z}{\partial t^2}+{\cal H}_z \Psi_z = 0,  \;\;\; \;\;\; {\cal H}_z = -\frac{\partial^2 }{\partial x^2} +V_z(r(x)),
\end{equation}
$x$  the tortoise radial coordinate, defined as
\begin{equation}\label{tortu}
x = r +2 M \ln \left| \frac{r-2M}{2M}\right|.
\end{equation}
Eq. (\ref{z})   admits separable solutions of the form  $e^{\pm i E^{1/2} t} \psi_E(x)$,
where ${\cal H}_z \psi_E = E \psi_E$.
The stability problem then reduces to finding out whether the ``Hamiltonian'' ${\cal H}_z$ admits
negative ``energy'' states or not. In the first case, $e^{\pm i E^{1/2} t} \psi_E(x)$ would grow exponentially
in time, as would do any solution of Zerilli's equation with initial data projecting non trivially on such a state.
If ${\cal H}_z$ is positive definite, the separate variable solutions are oscillatory in time,
and generic solutions of (\ref{z})  are then bounded for all $t$.
There is, however, a number of subtleties in the negative mass case, that were only recently settled
 \cite{dg2}. First note that  exterior region $r>2M$ of the $M>0$ Schwarzschild black hole gets mapped onto the real $x$ line
by (\ref{tortu}), whereas the $r>0$ region of interest in the negative mass case maps onto $x>0$, thus ${\cal H}_z$ is a ``quantum'' Hamiltonian on a half axis, with
 $V_z$ diverging
as $V_z \sim -1/(4x^2)$ as $ x \to 0^+$ \cite{dg2,ghi} (see \cite{reed,nc} for a detailed study of this problem.)
Zerilli's potential $V_z$ is smooth for $r>2M$ and positive $M$, whereas for $M<0$ has a second order pole
at $r=r_s>0$. The half space domain of (\ref{z}) for $M<0$, is related to the non global hyperbolicity
of the background spacetime.  This requires choosing a boundary condition at the $x=0=r$
singularity. As shown in Refs.1,4,5, there is a unique physically motivated choice that guarantees that
(i) first order corrections to  metric invariants do not diverge faster than their zeroth order piece as $x \to 0^+$,
guaranteeing that
the perturbed solution is globally valid, and (ii) the energy of the perturbation is finite\cite{ghi}.
The origin of the singularity at $r_s$ is different, and can be traced back to the definition of the Zerilli
function, which happens to be rational function  of the perturbed metric components, with an explicit singularity at this point\cite{dg,dg2}. 
As a consequence, Zerilli's function does not belong to $L^2({\mathbb R},dx)$,  ${\cal H}_z$ is not
a self adjoint operator on the space of Zerilli functions, and we do not know how to evolve initial data for perturbations. The way out
of this problem is provided by an appropriate intertwining operator\cite{price,dg2}, carefully constructed
so as to map Zerilli functions onto an $L^2$ space, and ${\cal H}_z$ onto a self adjoint Hamiltonian operator on this space. The unstable
mode found in Ref.1 gets mapped onto one of the (bound state) eigenfunctions of this Hamiltonian, and is therefore
excited by generic initial data supported away from the singularity. This completes the proof of instability of the negative
mass Schwarzschild black hole.\\
The unstable mode for $M<0$ Schwarzschild was recognized in Ref.11 to belong to the algebraic special modes (AS)
studied by Chandraskhar in the black hole context\cite{prs}. This hinted in the right direction to construct
unstable modes for the super-extremal Reissner-N\"ordstrom black hole, which also happened to be of the AS type\cite{dgp}.
Kerr spacetime breaks this pattern, as the AS modes do not satisfy appropriate boundary conditions  in the super extremal case\cite{dgp}. Perturbations of this spacetime are treated using Teukolsky equations\cite{teuko}, which describe the first order variation
$\psi$ of a null tetrad component  of the Weyl tensor, of spin weight $s= \pm2$. Separable solutions exist of the form
$\psi^{(s)} =
R_{\w,m,s}(r) S^m_{\w,s}(\theta) \exp(im\phi) \exp(-i\omega t)$. The equations they satisfy have the structure
\begin{equation} \label{ta}
{-1\over \sin\theta} {d \over d\theta}\left(\sin\theta {d S\over
d\theta}\right)+ {\cal A} (\theta, a \w, m, s) S = {E} S,
\end{equation}
\begin{equation}
\label{tr}
\Delta {d^2 R \over dr^2} +(s+1) {d\Delta\over dr} \;{dR\over dr} +
{\cal B} (r, \w, s, a, M) R = {E}  R,
\end{equation}
 This system is subtly linked: $S$ has to be regular on the sphere (these functions
on $S^2$ are called spin weighted spheroidal harmonics), $R$
has to satisfy appropriate  boundary conditions, and the eigenvalue $E$ is the same in (\ref{ta}) and (\ref{tr}).
Teukolsky equations were used  to prove that the exterior stationary region of  a Kerr black hole is stable\cite{stable},
and to find numerical evidence of an instability in the super-extremal case\cite{dgp}. A proof of the existence of the unstable modes
was recently found\cite{dgsv}, using the results  on the form of the spectrum of (\ref{ta}) for large
imaginary $\w=ik$ (corresponding to unstable modes $\sim \exp(kt)$)\cite{bertilong}
\begin{equation} \label{arrivaberti}
 E_{\ell}(a\w)|_{a\w=ik} = (2 \ell -3) k + {\cal O} (k^0), \;\;\; \text{ as }  k \to \infty, 
 \end{equation}
 together with its low frequency behavior
 \begin{equation} \label{abajoberti}
E_{\ell}(a\w)|_{a\w=0} = \ell(\ell+1) + {\cal O} (a \w)
\end{equation}
Here $\ell$ is the harmonic number, and the ones that are relevant for the perturbation problem are $\ell=2,3,...$.
The radial Teukolsky equation can be cast in Hamiltonian form ${\cal H} \psi = - \psi'' + V \psi = -E \psi$, the prime denoting a derivative with respect to an alternative radial coordinate. For $\w=ik$, $V=k^2 V_2(r)+kV_1(r)+V_0(r)$, with $V_1$ and  $V_2$  negative on an interval
$s_1(M)< r < s_2(M)<0$. This allows to prove that (minus) the lowest eigenvalue of this Hamiltonian behaves, for $k$ large enough, as
\begin{equation} \label{br}
 -\epsilon_o(k) \leq \;  \langle \psi | {\cal H} | \psi \rangle  <  k^2  \langle \psi |V_2 |\psi \rangle.
\end{equation}
The above equation, together with (\ref{arrivaberti}), (\ref{abajoberti}) and the bound $\epsilon_o(k=0^+)   <  \frac{15}{4}$
obtained from the global minimum of $V$ imply that, as we increase $k$, we always find
 a common $E$ eigenvalue in
(\ref{ta}) and (\ref{tr}), for any harmonic number and the fundamental radial mode\cite{dgsv}. Even more interesting is the
fact that, with minor changes, the same argument can be used to prove that, in the extreme ($J^2=M$) and
sub-extreme ($J^2<M$) black hole cases, the  stationary region beyond the inner horizon,  $r<r_i$,
 is  unstable\cite{dgsv}. The instability found in region III of Kerr space-time, and the fact that
the  Reissner-N\"orsdtrom charged black hole  also has a two horizon
structure with an inner static region $r<r_i$, triggers the question of
whether of not this inner static region is stable.
Preliminary work indicates
that this region is unstable\cite{futuro}.\\

\end{document}